\documentclass[useAMS,usegraphicx,usenatbib]{mn2e}

 \long\def\symbolfootnote[#1]#2{\begingroup%
 \def\thefootnote{\fnsymbol{footnote}}\footnote[#1]{#2}\endgroup}

\usepackage{subfigure}
\usepackage{graphicx}

\title[MUSE observations of PKS1614+051]{Dissecting the complex environment of a distant quasar with MUSE}
\author[K. Husband et al.]{K. Husband$^1$, M.~N. Bremer$^1$\thanks{Email: m.bremer@bristol.ac.uk}, E.~R. Stanway$^2$, M.~D. Lehnert$^3$ \\
$^1$H.H.~Wills Physics Laboratory, University of Bristol, Tyndall Avenue, Bristol, BS8 1TL, UK. \\
$^2$Department of Physics, University of Warwick, Gibbet Hill Road, Coventry CV4 7AL, UK. \\
$^3$Institut d'Astrophysique de Paris, UMR 7095, CNRS, Universit\'{e} Pierre et Marie Curie, 98 bis Boulevard Arago, F-75014 Paris, France. \\
}

\begin{document}
\maketitle
\begin{abstract}
High redshift quasars can be used to trace the early growth of massive
galaxies and may be triggered by galaxy-galaxy interactions. We
present MUSE science verification data on one such interacting system
consisting of the  well-studied $z=3.2$ PKS1614+051 quasar, its
AGN companion galaxy and bridge of material radiating in Ly$\alpha$
between the quasar and its companion. We find a total of four
companion galaxies (at least two galaxies are new discoveries),
three of which reside within the likely virial radius of the quasar
host, suggesting that the system will evolve into a massive elliptical
galaxy by the present day.  The MUSE data are of sufficient quality
  to split the extended Ly$\alpha$ emission line into narrow velocity
  channels. In these the gas can be seen extending towards each of the
  three neighbouring galaxies suggesting that the emission-line gas
  originates in a gravitational interaction between the galaxies and
  the quasar host. The photoionization source of this gas is less
  clear but is probably dominated by the two AGN.  The quasar's
Ly$\alpha$  emission spectrum is double-peaked, likely due to
  absorbing neutral material at the quasar's systemic redshift with a
low column density as no damping wings are present. The spectral
  profiles of the AGN and bridge's Ly$\alpha$ emission are also
consistent with absorption at the same redshift indicating this
neutral material may extend over $>$ 50 kpc. The fact that the neutral
material is seen in the line of sight to the quasar and transverse to
it, and the fact that we see the quasar and it also illuminates the
emission-line bridge, suggests the quasar radiates isotropically and
any obscuring torus is small. These results demonstrate the power of
MUSE for investigating the dynamics of interacting systems at high
redshift.

\end{abstract}


\section{Introduction}
The growth of galaxies and galaxy clusters
can be probed using high redshift quasars to pinpoint regions of early
cluster formation \citep[e.g.][]{Turner91,Springel05,DiMatteo08}. The
rapid buildup in mass of the central supermassive black holes and host
galaxies of quasars implies that they are hosted by massive dark
matter haloes - themselves the likely sites of early galaxy
clustering. Thus distant quasars can be used to explore the early
interactions that may trigger quasar activity and lead to the growth
of massive galaxies that eventually exist at the centres of
present-day clustering. Observational evidence of galaxy clustering
around at least a subset of high redshift quasars is now becoming
apparent \citep[e.g.][]{Toshikawa12,Husband13,McGreer14}.

One of the earliest identified systems that showed evidence of
interaction and clustering at high redshift is that of PKS1614+051,
consisting of a $z=3.2$ radio-loud quasar and companion galaxy
connected by an extended bridge of ionized gas. The emission-line
companion galaxy is itself an AGN \citep{Djorgovski87, Bremer95} and
the bridge of gas is radiating strongly in Ly$\alpha$
\citep{Hu&Cowie87}. While not unique in having clearly extended
Ly$\alpha$, the combination of quasar, AGN and extended Ly$\alpha$
emission is a rare opportunity to explore the connection between a
high redshift quasar and its environment. First discovered by
\citet{Djorgovski85}, the brightness of the companion galaxy
(significantly brighter than typical Lyman break galaxies and
Ly$\alpha$ emitters at $z\sim3$) and the Ly$\alpha$ bridge allowed the
system to be studied with 4-m telescopes in the 1980s and 1990s. Given
the companion galaxy, some authors \citep[e.g.][]{Hu&Cowie87} suggest
it should lie in an overdense (or protocluster) environment that will
evolve into a galaxy cluster by the present day.

With the advent of MUSE \citep[Multi-Object Spectroscopic
  Explorer;][]{Bacon12} - a 1 x 1 arcmin$^{2}$ field integral field
unit (IFU) on the VLT (Very Large Telescope) - a more detailed study
of this region is now possible. The large IFU enables continuum and
emission line morphologies to be acquired as well as information about
the dynamics and ionization state of the gas and galaxies, without
which interpretation of the complex interaction of quasars with their
environments is highly challenging. Often the close environment of a
quasar is studied using rare, spatially coincident background quasars
to probe individual sight lines through the halo containing the
quasar. Here we show that we can probe numerous lines of sight around
a quasar using the IFU capabilities of MUSE and the Ly$\alpha$ halo of
the quasar. Although already well-studied, the MUSE science
verification data presented in this paper provides new information on
the nature of this system.

A $\Lambda$CDM cosmology with $H_{0}=70$ km s$^{-1}$ Mpc$^{-1}$,
$\Omega_{M}=$ 0.29 and $\Omega_{\Lambda}=$ 0.71 \citep{Bennett14}
  is used throughout.

\section{Observations}
The data were taken as part of Science Verification with VLT/MUSE
between 18 and 20th August 2014. The 1 x 1 arcmin$^2$ field of view
was first centered 15 arcsec west of PKS1614+051 and then 15 arcsec to
the east. Four 880s exposures, rotated 90 deg between each, were taken
at each position, giving a total exposure time of 1.96 h and
spectroscopic coverage of all galaxies within 1.28 x 1 arcmin$^2$ over
a wavelength range of 4750-9350 \AA\ . This arrangement meant that the
central $\sim$30 arcsec wide North-South strip of the eventual image
accrued twice the exposure time of the $\sim$15 arcsec wide eastern
and western strips.

\begin{figure*}
\center
\includegraphics[width=2.\columnwidth]{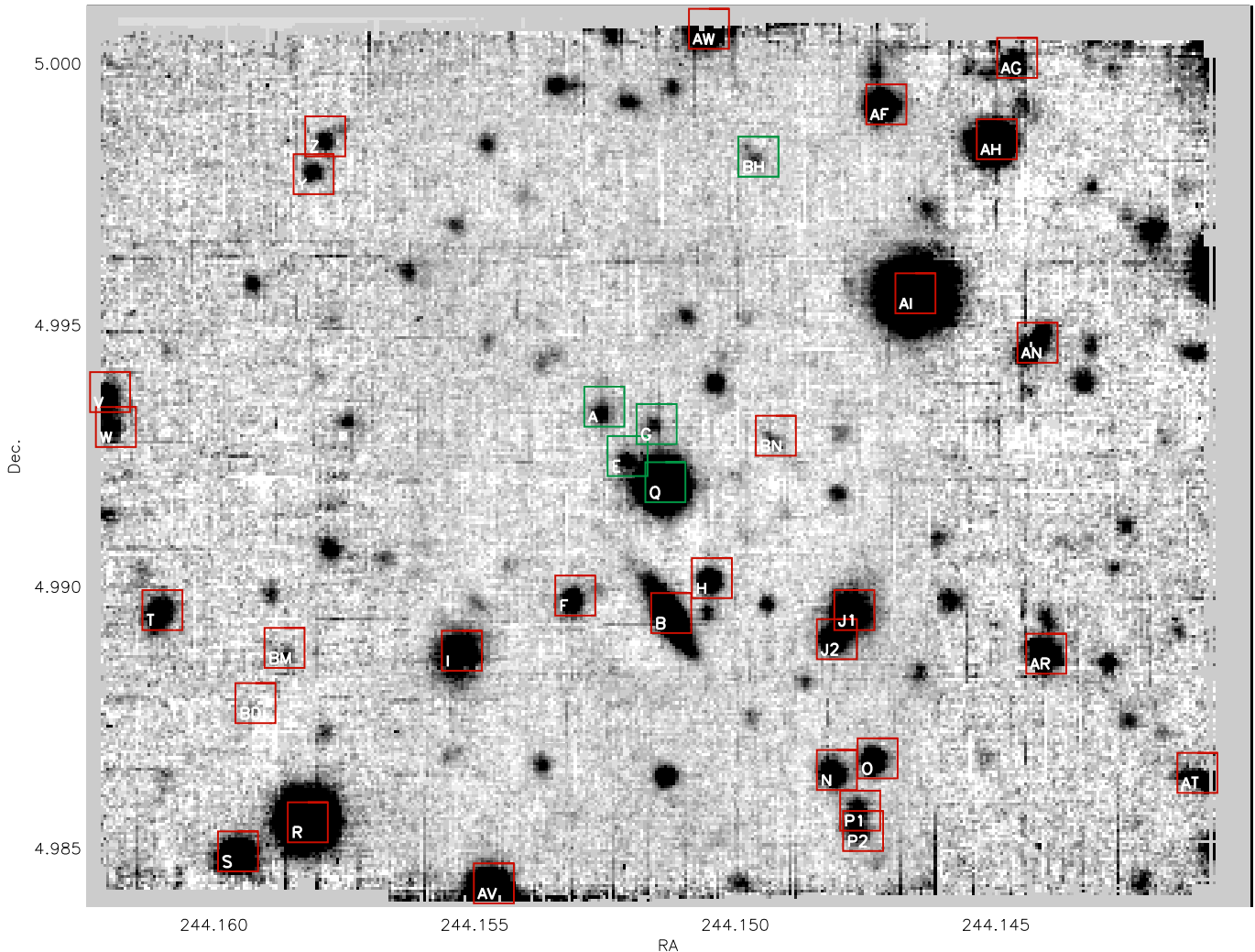}
\vspace*{0.2in}
\caption{A white light image (4750-9350 \AA) of the MUSE field of view
  highlighting the galaxies whose redshift could be determined as well
  as the stars that could be identified. Galaxies that are at the
  same redshift as the central quasar (Q) are highlighted in green (A,
  E, G and BH), whilst those at lower redshift are in red. The
  labeling for the central objects is the same as in
  \citet{Hu&Cowie87} with Q marking the quasar. All the details for
  these objects are in Table \ref{allresults}.  }
\label{superimage}
\end{figure*}

The MUSE Data Reduction Software (version 0.18.5) was used for the
first reduction of the data - bias subtracting, flat fielding and
calibrating the data, and aligning and constructing three dimensional
data cubes. An IDL script was written to achieve a better sky
subtraction than the early MUSE pipeline by fitting and subtracting a
2nd order polynomial to each spatial row (along the long axis of the
IFUs) of each wavelength slice of the resulting data cubes for each of
the eight exposure in order to remove the sky contribution. The
resulting eight data cubes (one for each of the four rotations at two
separate pointings) were mean-combined with cosmic ray rejection
(sigma clipping) and the median of the entire frame at each wavelength
bin subtracted to remove any remaining residual from the sky
subtraction. Images were constructed from the data cubes by collapsing
the spectral axis over defined wavelength ranges. A further median
subtraction was done to each row of each individual IFU for every
exposure and then these were combined with a bad pixel mask and sigma
clipping to produce the final white-light image (see
Fig.~\ref{superimage}).

The RMS noise in the final data cube is $1\times10^{-19}$ erg s$^{-1}$
cm$^{−2}$ per 1.25 \AA\ wavelength bin, at 5150 \AA\ and in the center
of the data cube. The RMS noise is a factor of $\sqrt{2}$ worse in the
two 15 arcsec wide strips on the east and west of the image where the
total exposure time is half that of the middle and the noise is up to
50 per cent lower in the middle of the wavelength range around 7000
\AA. The 2$\sigma$ depth, within 2 arcsec diameter apertures, of the
white light, full range (4750-9350 \AA) stacked image is 26.4 AB. The
FWHM velocity resolution is measured from the sky lines to be 2.7
\AA\ or equivalently 150 km s$^{-1}$ at $z=3.215$ and the spatial FWHM
of the final white light image is 0.9 arcsec. The pixel scale of MUSE
is 0.2 arcsec (pixel)$^{-1}$.

The final white light image is shown in Fig.~\ref{superimage}. We use
SExtractor \citep{Bertin96} followed by a visual check to identify 119
objects in our white-light image, of which 60 are brighter than $R=25$
AB (see Table \ref{allresults}). We also checked the entire data cube
for emission lines but only one additional object was found. This is
unsurprising as any emission line only object would have an extremely
high equivalent width and likely a very red or very blue spectrum in
order for the continuum to remain undetected. The only other object
with undetected continuum is the Ly$\alpha$ bridge indicating it has a
very large rest-frame equivalent width ($>$60 \AA\ using the 2$\sigma$
limit on the spectrum in the wavelength range 5090-5190 \AA\ near the
Ly$\alpha$ emission or $>$180 \AA\ using the 2$\sigma$ limit on the
full range white light image shown in Fig.~\ref{superimage}) that is
unlikely to be powered by embedded star formation in the bridge
\citep[the maximum equivalent width from star forming regions is 240
  \AA;][]{Hennawi13,Charlot93}. Analysis of both emission lines,
absorption lines and comparison with Sloan Digital Sky Survey (SDSS)
galaxy and stellar templates results in redshifts for 34 galaxies
(Table \ref{allresults}), of which four are at the redshift of the
quasar (within $\Delta z=0.003$). We are sensitive to Ly$\alpha$ flux
densities of greater than $2\times10^{-19}$ erg s$^{-1}$ cm$^{-2}$
\AA$^{-1}$ at the redshift of the quasar, determined by collapsing a
100 \AA\ wide section of the data cube around 5120 \AA\ over the
wavelength axis. In addition, we found no convincing Lyman break
galaxies (LBGs) without Ly$\alpha$ emission when we compared stacked
images above and below potential Ly$\alpha$ breaks at 20
\AA\ intervals over the full wavelength range.

 \begin{table}
   \centering
    \begin{tabular}[t]{c c c c c c}
    \hline
    ID & RA &  Dec & Redshift  & Classification \\
    \hline
    Q & 16:16:37.56 & 04:59:32.1 & 3.212 & Quasar \\
    A & 16:16:37.83 & 04:59:37.1 & 3.214 & LAE/NLAGN$^1$ \\
    B & 16:16:37.52 & 04:59:22.9 & 0.507 & Spiral galaxy \\
    E & 16:16:37.72 & 04:59:33.8 & 3.215 & LAE \\
    F & 16:16:37.97 & 04:59:24.0 & 0.67  & Galaxy  \\
    G & 16:16:37.59 & 04:59:36.2 & 3.215 & LAE     \\
    H & 16:16:37.34 & 04:59:25.4 & 0.000 & M star  \\
    I & 16:16:38.49 & 04:59:20.5 & 0.143 & Galaxy \\
    J1& 16:16:36.69 & 04:59:23.3 & 0.53  & Early-type galaxy \\
    J2& 16:16:36.77 & 04:59:21.3 & 0.615 & Luminous red galaxy \\
    L & 16:16:38.11 & 04:59:12.8 & 3.098 & LBG/LAE \\
    N & 16:16:36.77 & 04:59:12.3 & 1.089 & Galaxy or \\
      &             &            &       & Interacting galaxies \\
    O & 16:16:36.58 & 04:59:13.1 & 0.000 & M Star \\
    P1& 16:16:36.66 & 04:59:09.5 & 0.608 & Galaxy \\
    P2& 16:16:36.65 & 04:59:08.1 & 0.608 & Galaxy \\
    R & 16:16:39.20 & 04:59:08.7 & 0.000 & Star \\
    S & 16:16:39.52 & 04:59:06.7 & 0.402 & Galaxy \\
    T & 16:16:39.87 & 04:59:23.3 & 0.905 & Galaxy \\
    V & 16:16:40.11 & 04:59:38.3 & 0.71  & Galaxy \\
    W & 16:16:40.08 & 04:59:35.9 & 0.16  & Galaxy \\
    Y & 16:16:39.17 & 04:59:53.3 & 1.470 & Galaxy \\
    Z & 16:16:39.12 & 04:59:55.9 & 0.000 & M star \\
    AF& 16:16:36.54 & 04:59:58.1 & 0.000 & M star \\
    AG& 16:16:35.94 & 05:00:01.3 & 1.213 & Galaxy \\
    AH& 16:16:36.03 & 04:59:55.7 & 0.833 & Galaxy \\
    AI& 16:16:36.40 & 04:59:45.1 & 0.000 & Star  \\
    AR& 16:16:35.80 & 04:59:20.3 & 0.629 & Galaxy \\
    AT& 16:16:35.11 & 04:59:12.1 & 0.549 & Spiral galaxy \\
    AV& 16:16:38.34 & 04:59:04.5 & 0.000 & Star \\
    AW& 16:16:37.35 & 05:00:03.3 & 0.000 & M star \\
    BD& 16:16:39.44 & 04:59:16.9 & 0.143 & Galaxy \\
    BH& 16:16:37.13 & 04:59:54.5 & 3.217 & LAE     \\
    BM& 16:16:39.31 & 04:59:20.7 & 0.853 & Galaxy \\
    BN& 16:16:37.05 & 04:59:35.3 & 1.050 & Galaxy \\
    \hline
    \end{tabular}
    \caption{A summary of the redshifts obtained. The ID is the same
      as in Fig.~\ref{superimage} and is an extension of the labeling
      scheme of \citet{Hu&Cowie87}. We do not find galaxies at the
      positions of their C and D objects. $^1$\citet{Bremer95} }
    \label{allresults}
\end{table}

\section{Results}
The capabilities of MUSE are clearly demonstrated in this data - with
just two hours of observations we can identify galaxies at the same
redshift as the quasar across the 1 arcmin$^2$ field of view and probe
in detail the region close to the quasar containing PKS1614, its AGN
companion, the emission line bridge and any other potential companion
galaxies.

\subsection{Galaxy Environment of PKS1614}
\begin{figure}
\centering
\includegraphics[width=0.95\columnwidth]{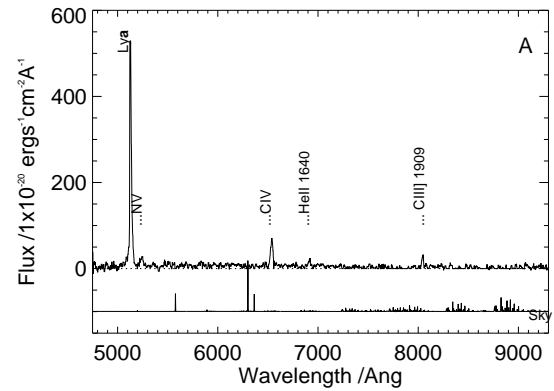}
\caption{The extracted 1D spectrum of companion A with the sky lines
  plotted underneath. }
\label{spectra_A}
\end{figure}

\begin{figure}
\centering
\includegraphics[width=0.95\columnwidth]{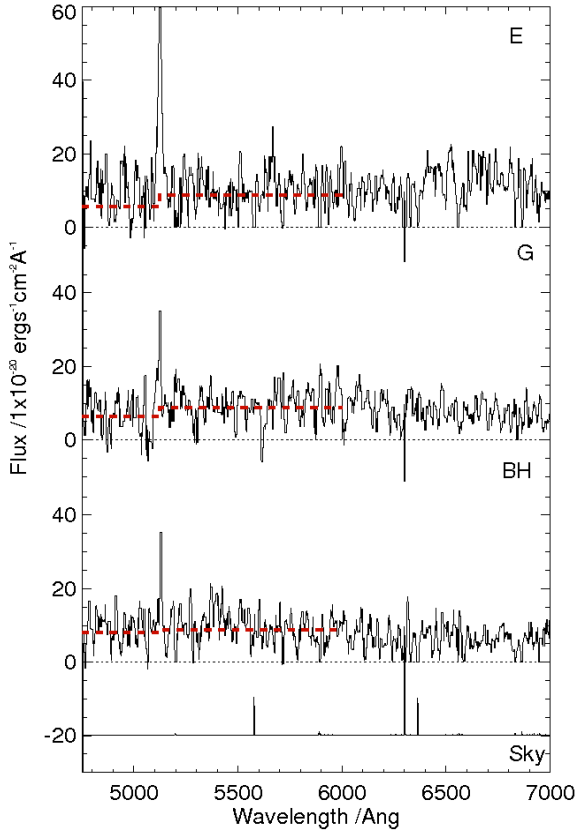}
\caption{The extracted 1D spectrum of companion E, G and BH with the
  sky lines plotted underneath. The dashed line shows the median of
  the spectrum either side of the expected break. Areas of the
  spectrum affected by skylines have been set to zero.  }
\label{spectra_companions}
\end{figure}

The PKS1614+051 quasar has at least four companion galaxies: the
previously known narrow line AGN A, galaxy E - hitherto identified
as a candidate $z=3.2$ Ly$\alpha$ emitter (LAE) based on narrow-band
imaging \citep{Hu&Cowie87,Hu91} - and two new LAEs/LBGs: G and BH. For
these newly identified sources, we extend the galaxy labeling scheme
of \citet{Hu&Cowie87} but we do not find galaxies at the positions of
their C and D objects. The spectra of these companion galaxies are
shown in more detail in Fig.~\ref{spectra_A} \&
\ref{spectra_companions}.

Three of the companion galaxies (A, E and G) are close enough to be
interacting with the quasar as defined by \citet{Ellison08} and are
within the expected virial radius of the quasar's host dark matter
halo suggesting they will eventually merge with the quasar host to
form a massive elliptical galaxy (perhaps at the centre of a group or
cluster) by the present day. Although the spectrum of E may be
affected by scattered light from the quasar, its broad-band photometry
is consistent with a Lyman break and its Ly$\alpha$ line is
significantly narrower than the quasar's, suggesting it is a separate
$z=3.2$ galaxy. E is unlikely to be driven solely by fluorescence from
the quasar  \citep[e.g.][]{Cantalupo05,Kollmeier10} as the presence of
continuum emission suggests star formation can create the Ly$\alpha$
emission (the rest-frame equivalent width is only $\sim$15 \AA). The
companion galaxies are all cleanly separated from the quasar
demonstrating the excellent spatial sampling of the MUSE data.

In total, we confirm five galaxies including the quasar host within an
area of 20 by 25 arcsec by $\Delta z=0.003$ or a maximum co-moving
volume of 1.4 co-moving Mpc$^{3}$, ignoring peculiar velocities
i.e.~assuming the difference in redshift represents the Hubble
expansion.

 \begin{table*}
 \centering
 \begin{tabular}[t]{c c c c c c c c c}
 \hline
 ID & z$_{peak}$ & z$_{sys}$ & z$_{fit}$ & Ly$\alpha$ flux & Ly$\alpha$ Luminosity & Ly$\alpha$ Vel.~Dispersion & R Abs.~Mag. \\
 & & & & /10$^{-20}$ erg s$^{-1}$ cm$^{-2}$ & /10$^{43}$ erg s$^{-1}$ & /km s$^{-1}$ & /mag \\
 \hline
 Q   & 3.215 & 3.212 & 3.215 & 616720 & 58.8   & 920 & -28.1 \\
 A   & 3.215 & 3.214 & 3.215 & 33260 & 3.18   & 400 & -23.0 \\
 E   & 3.215 & na & 3.215 &  3960 & 0.035  & 260 & -23.2 \\
 G   & 3.215 & na & 3.211 &  1080 & 0.103 & 190 & -22.7 \\
 BH  & 3.217 & na & 3.217 &   773 & 0.074 & 130 & -22.9 \\
 Bridge & 3.215 & na & 3.214 & 10940 & 1.989 & 250 & -21.0 \\
 \hline
 \end{tabular}
 \caption{The properties of the companion galaxies relative to the
   quasar. The Ly$\alpha$ flux is determined from the area under a
   Gaussian fit to the line with the continuum removed in a 2 arcsec
   aperture. The properties for the bridge are for the entire bridge
   structure once the contribution from the quasar and companion A are
   removed, except for the velocity dispersion which is the value for
   the centre of the bridge.}
 \label{results}
\end{table*}

\subsection{Gaseous Environment}
\begin{figure*}
\center
\subfigure{\includegraphics[width=0.89\columnwidth]{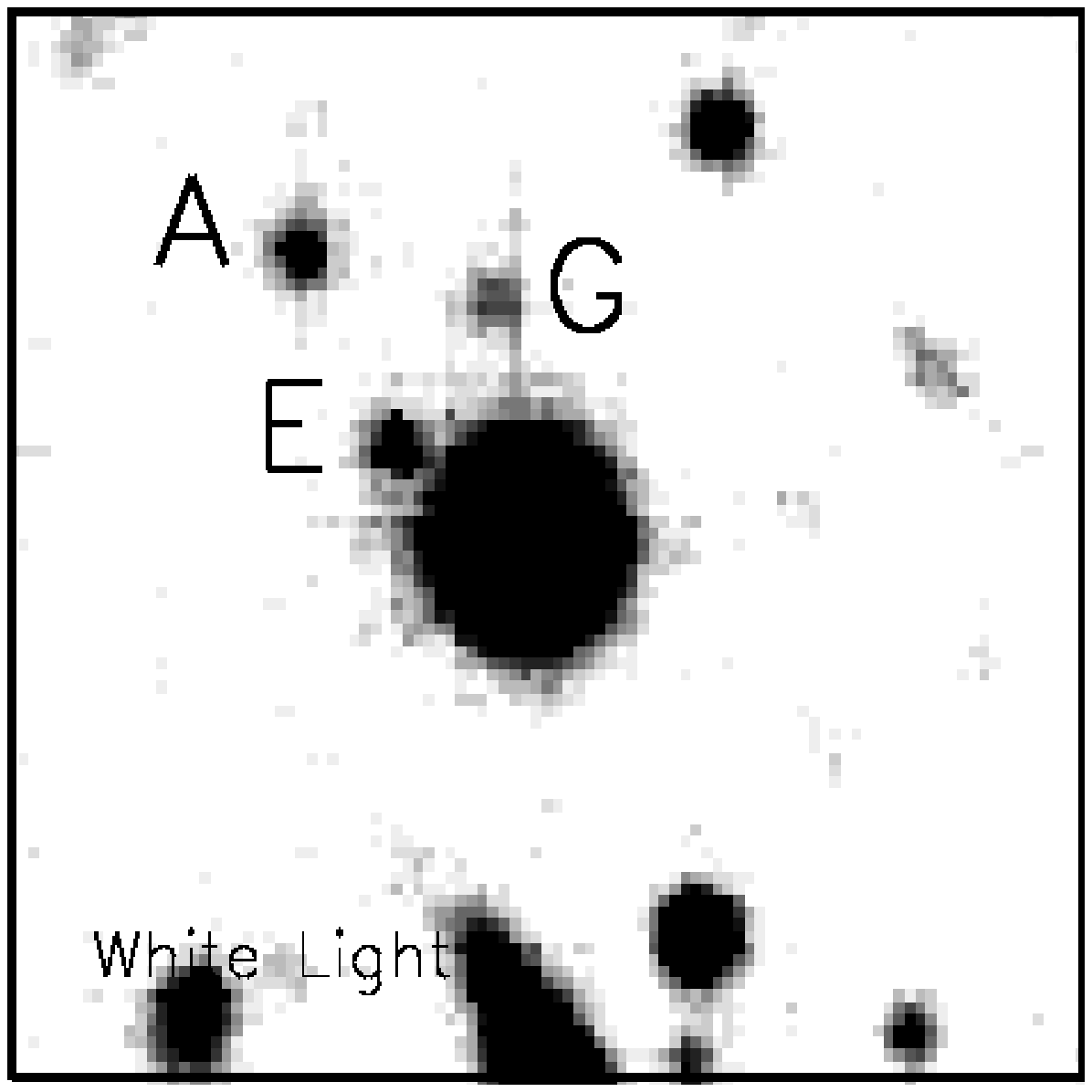}}
\subfigure{\includegraphics[width=0.89\columnwidth]{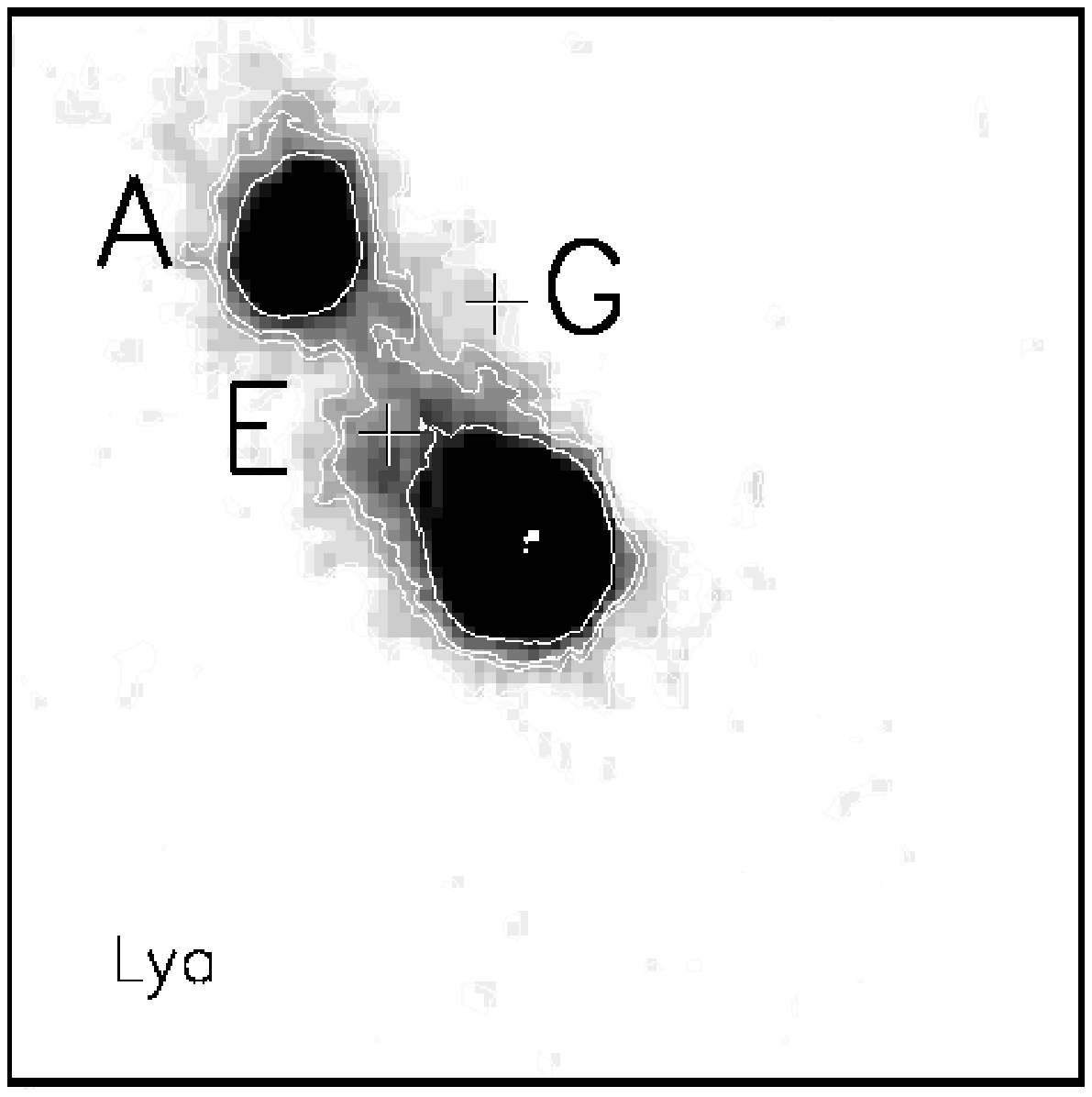}}
\subfigure{\includegraphics[width=0.89\columnwidth]{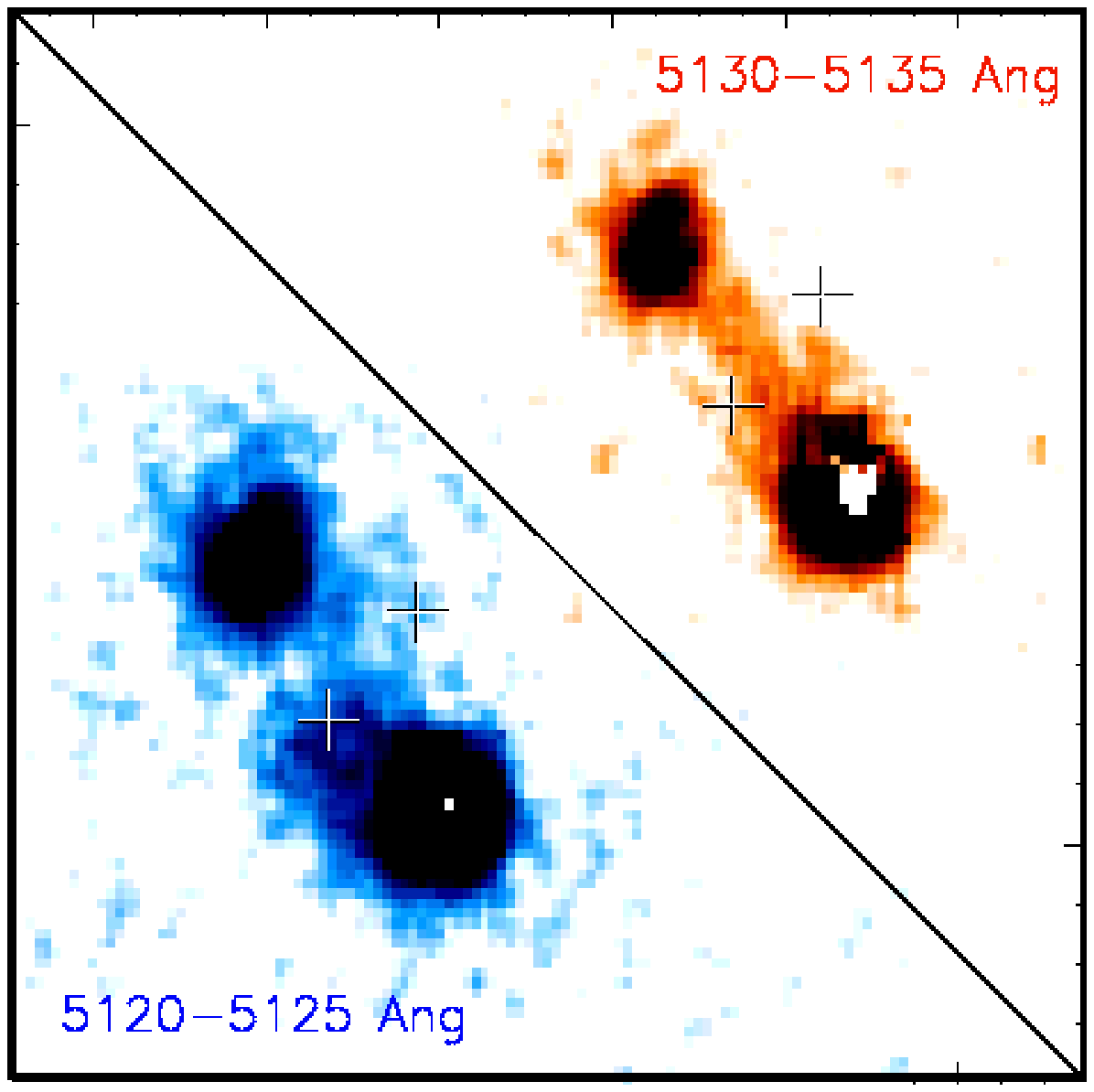}}
\subfigure{\includegraphics[width=0.89\columnwidth]{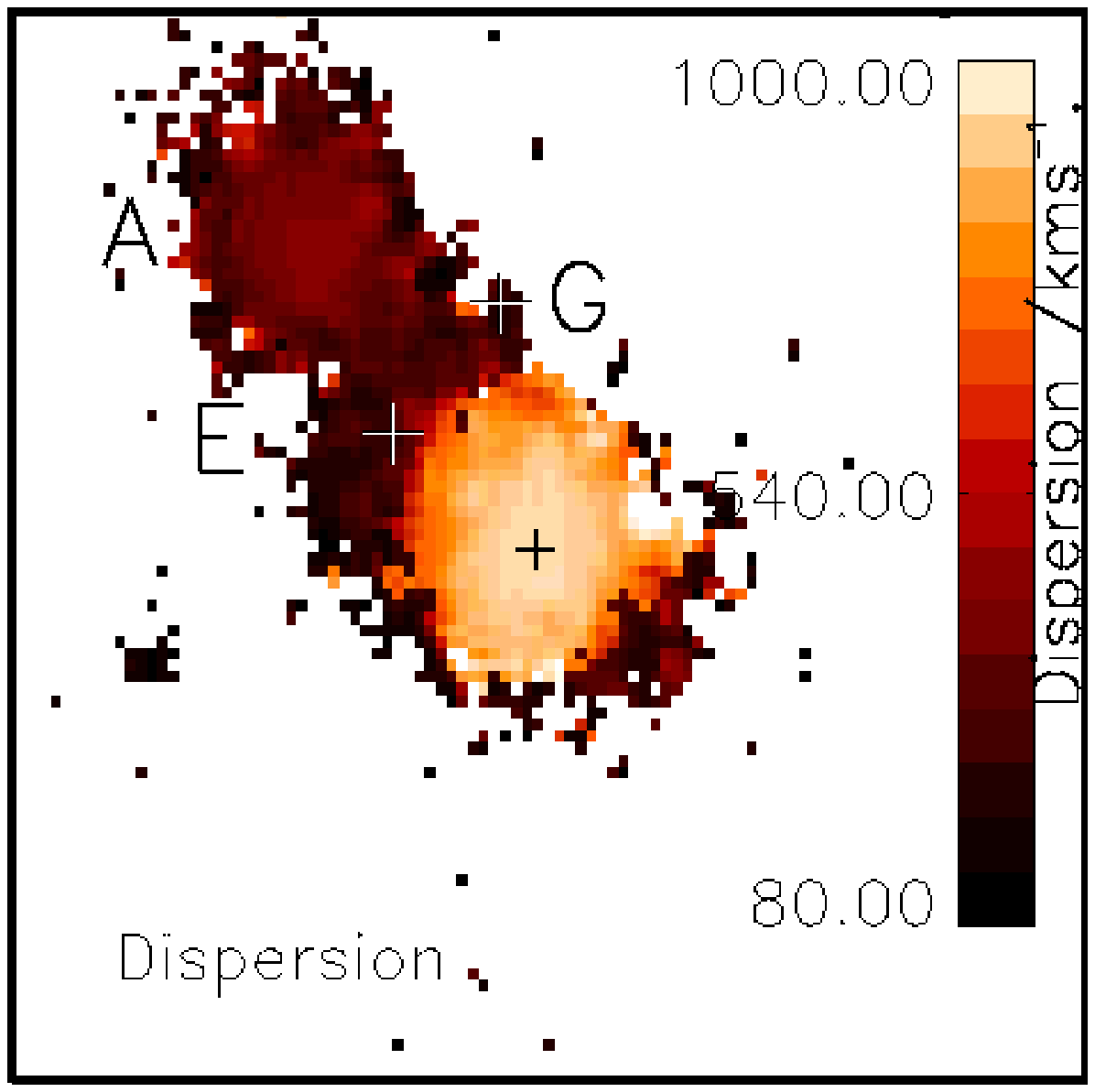}}
\caption{{\bf a.} White light image of the centre of the MUSE field of
  view (20 by 18 arcsec) clearly showing galaxies E and G. {\bf
      b.}  The same area in the narrow band (5120-5135 \AA)
    encompassing Ly$\alpha$. Some of the Ly$\alpha$ emission from the
    quasar and A has also been subtracted using a PSF model in order
    to show the extended emission more clearly. The contours trace the
    Ly$\alpha$ showing that both the quasar and A are clearly extended
    in Ly$\alpha$ and the emission extends beyond A to the North. The
    crosses mark the positions of galaxies E and G. The surface
    brightness limit is $\sim 4\times10^{-17}$ erg s$^{-1}$ cm$^{-2}$
    arcsec$^{-2}$ and the contours indicate surface brightness of (4,
    5, 9, 11 and 20) $\times10^{-17}$ erg s$^{-1}$ cm$^{-2}$
    arcsec$^{-2}$. {\bf c.}  The Ly$\alpha$ line is split into the
    blue-most (5120-5125 \AA) and red-most (5130-5135 \AA) velocity
    components. In these narrower components the gas is seen to
    extended between the quasar and A and also towards galaxies E and
    G (marked by crosses). {\bf d.}  The velocity dispersion of the
  Ly$\alpha$ line from the same Gaussian fits to each spatial
  pixel. This shows that the quasar has a velocity dispersion of 920
  km s$^{-1}$ whilst companion A has a velocity dispersion of 400 km
  s$^{-1}$. The high values are what is expected for active
  galaxies. Companion E and the material in the bridge have much lower
  velocity dispersions, around the velocity resolution of MUSE. }
\label{peak}
\end{figure*}

 We have obtained a detailed, high signal-to-noise spectrum of the
  quasar, companion A and, for the first time, the emission line
  bridge.  Fig.~\ref{peak} shows the distribution of gas in the
  central region around the quasar. The emission line bridge extends
  over 50 physical kpc from A to the quasar and slightly beyond the
  quasar in the other direction (Fig.~\ref{peak}b), with a width of
  around 10 kpc. Fig.~\ref{peak}c shows the red-most and blue-most (5
  \AA\ wide or 300 km s$^{-1}$) components of the Ly$\alpha$ line,
  showing how the high surface brightness emission connects the quasar
  to companion A and also extends towards the two other nearby
  galaxies (E and G), denoted by crosses.  Fig.~\ref{peak}d shows the
  dispersion of the Ly$\alpha$ line as a function of position, found
  from fitting a Gaussian to the Ly$\alpha$ line in each spatial
  pixel. The Ly$\alpha$ emission from the bridge has a relatively low
  dispersion, as does the Ly$\alpha$ emission from companions E and G,
  whilst the emission from the quasar and companion A have larger
  velocity dispersions as expected for active galaxies (see Table
  \ref{results}).

\begin{figure}
\centering
\includegraphics[width=0.95\columnwidth]{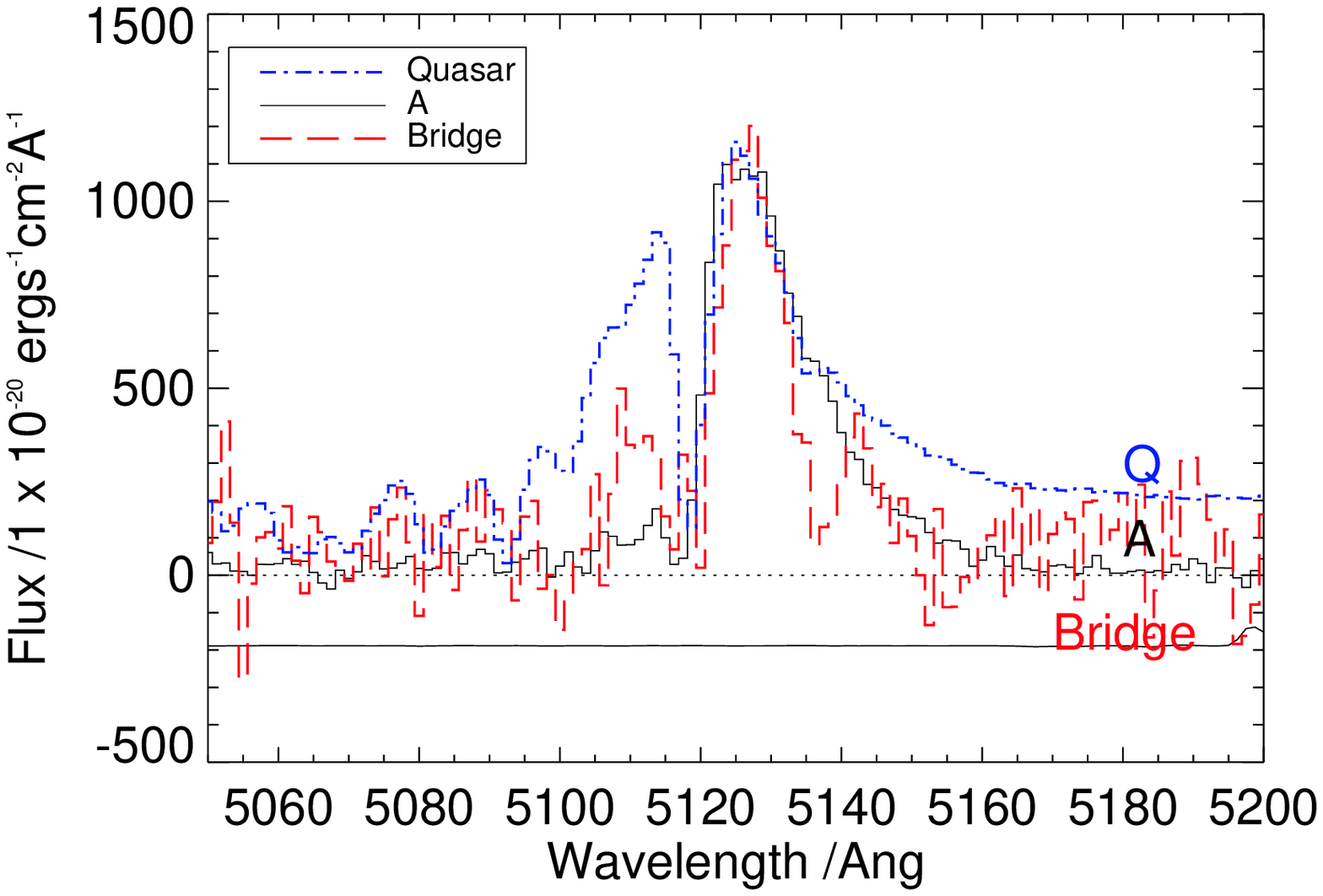}
\caption{A zoom in on the extracted 1D spectrum of the quasar,
  companion A and the bridge showing the Ly$\alpha$ emission
  profile. The flux of the bridge and quasar have been scaled to that
  of companion A. There is strong absorption at 5120 \AA\ that is
  Ly$\alpha$ at the quasar's systemic redshift. All three objects show
  a sharp drop at this wavelength. The bridge appears to recover flux
  short-ward of the drop, as may companion A but note that the signal
  to noise on the spectrum of A is much higher than the signal to
  noise of the bridge. }
\label{spectra_Qzoom}
\end{figure}

 The spectrum of the quasar's Ly$\alpha$ emission is strongly
  double-peaked with the central gap having a rest-frame equivalent
  width of 1.1 \AA\ and the flux declining to almost zero at the
  central wavelength of the line. If this is due to an absorption
  system the velocity width, b, is 180$\pm$15 km s$^{-1}$ and this
  implies a low column density of ($1.9\pm 0.4$) x 10$^{13} $
  cm$^{-2}$ using a curve of growth analysis and assuming a covering
  fraction of 100 per cent, as expected for an unsaturated absorption
  profile without damping wings. This potential absorbing material is
  at the systemic redshift of the quasar (as derived from other lines
  e.g.~HeII, CIV) suggesting that the emission arises from neutral
  material in or around the quasar. The spectra of A and the bridge
  are also consistent with being affected by the same absorbing
  material. Their spectral profiles display a similar sharp cutoff on
  the blue-ward edge as the quasar's Ly$\alpha$ profile
  (Fig.~\ref{spectra_Qzoom}), although the underlying Ly$\alpha$
  emission lines in A and the bridge are narrower and appear to peak at
  slightly longer wavelengths than the quasar Ly$\alpha$
  emission line. 

 We see no other emission lines arising from the bridge, placing a
  constraint on the CIV/Ly$\alpha$ and HeII/Ly$\alpha$ flux ratios of
  $<0.03$ and $<0.06$ (3$\sigma$), significantly below the ratios for
  the quasar and companion A where CIV and HeII was detected. This
  implies a lower ionization parameter for the bridge than that
  typically seen from an AGN (see discussion section below).

\section{Discussion}
\subsection{Protocluster or Protogalaxy Environment}
PKS1614+051 lies in a galaxy overdensity of at least five galaxies
(including itself) within a volume of no more than 1.4 co-moving
Mpc$^{3}$, demonstrating that MUSE is clearly capable of identifying
clustering and potential protoclusters at $z\sim3$. At this redshift,
MUSE allows for little spectral coverage below the 1216
\AA\ Ly$\alpha$ break. Given the relative weakness of this break at
$z=3$ when compared to higher redshifts, and the low signal to noise
obtained on the continuum of faint sources at short wavelengths, it
will be easier to use MUSE to search for Lyman break galaxies at
higher redshifts than that probed here.

The majority (3/4) of the companion galaxies are within a 50 kpc of
the quasar, and hence well within its expected virial radius
\citep[$\sim$ 280 kpc;][]{Cantalupo14}, suggesting that they will
merge with the quasar to form a massive elliptical galaxy by
$z=0$.  If this is the progenitor of the massive galaxies seen at
  $z\sim1.5$ then the merger needs to be completed within $\sim$ 2.5
  Gyr.

At this redshift we expect protoclusters to extend over radii of 2-5
arcmin \citep{Chiang13}. As MUSE only probes 1 arcmin$^2$ we can draw
limited conclusions about the wider protocluster environment of
PKS1614. Nevertheless, with one companion galaxy outside its virial
radius and three within, the quasar is highly likely to reside in an
overdense environment likely to evolve into a group or cluster.  We
note that with another eight similar pointings - in total 18 hours of
MUSE time - the environment of the system could be probed in detail on
the scale of an ACS or WFC3 HST image, better matched to the expected
scale of overdense regions.

\subsection{Gaseous Interactions}
 High redshift radio loud quasars and galaxies are often observed
  to have extended Ly$\alpha$ haloes surrounding them, spanning
  10s-100s of kpc
  \citep[e.g.][]{Heckman91a,Christensen06,Villar-Martin07} with line
  widths of 100-1500 km s$^{-1}$ \citep{Heckman91b}. The
  photoionization of the gas is often attributed to radiation escaping
  along the radio axis of the galaxy, although fainter extended
  Ly$\alpha$ haloes are seen around compact radio galaxies
  \citep[e.g.][]{vanOjik97} and around some radio-quiet quasars
  \citep[e.g.][]{Christensen06,Herenz15}. Many of the extended
  emission line regions around radio galaxies ($\sim$ 60 per cent) and
  in particular those around the smallest ($<50$ kpc sized) radio
  sources such as PKS1614+051, have associated HI absorption
  \citep{vanOjik97}. Indeed $\sim60$\% of the 74 $z\sim2$ quasars
  studied by \citet{Prochaska13b} were found to have nearby ($< 200$
  kpc) neutral material absorbing photons from a spatially coincident
  background quasar. Assuming the quasar is at the centre of a similar
  large-scale diffuse, high filling factor neutral halo, the picture
  we favour to explain the observed distribution of gas around
  PKS1614+051 is as follows: Embedded within this diffuse halo are
  also the companion galaxies A, E and G and the bridge of denser,
  lower filling factor emission-line gas.  This gas phase originates
  from material extracted from the companions and quasar host through
  mutual gravitational interactions and is ionized by AGN and/or
  stellar emission.  Below we discuss the evidence in favour of this
  picture.

 The gas in the bridge is aligned between the quasar and companion
  A, and may extend slightly beyond A and the quasar along the same
  axis. Fig.~\ref{peak}c shows the distribution of the red-most and
  blue-most component of the Ly$\alpha$ bridge gas. There is a
  comparatively narrow stream of gas connecting the quasar to
  companion A, with a wavelength similar to the Ly$\alpha$ peak
  wavelength of A, suggesting the two galaxies are gravitationally
  interacting. The bridge also seems to extend towards galaxies E and
  G (marked by crosses), giving it its broad appearance in the
  full-width Ly$\alpha$ image and indicating that these galaxies may
  also be interacting with the bridge and/or quasar.
  
 Unlike many powerful high redshift radio sources, in this case
  there is no obvious spatial relationship between the radio emission
  and the extended emission-line region. Radio observations of the
  quasar indicate that it is a gigahertz peaked source with the bulk
  of the radio emission arising from a milli-arcsec scale region. The
  source's radio jet originates from the quasar in a south-westerly
  direction before turning through 90 degrees to the north-west again
  on sub arcsec scales \citep{Djorgovski87,Orienti06} and hence the
  extended emission line region is probably aligned with the
  counter-jet axis, albeit on scales two or more orders of magnitude
  larger than that of the radio emission. However, the lack of
  similarly extended emission in the south-westerly direction suggest
  the radio-jet is not the primary cause of the emission-line
  bridge. 

\begin{figure}
\center
\includegraphics[width=0.95\columnwidth]{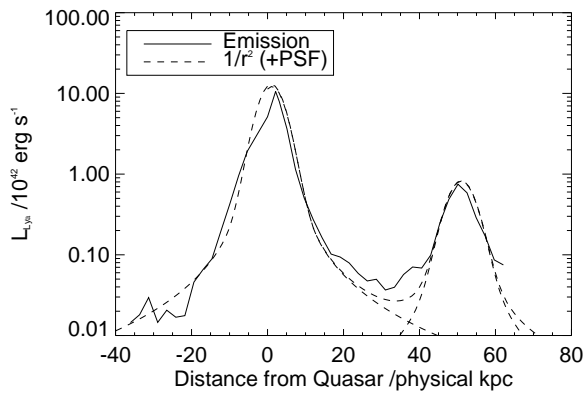}
\caption{A comparison of the Ly$\alpha$ luminosity along a line
  through the quasar and companion A (taking a cross section along the
  axis of the bridge).  The line that seems to fit the data is a
    1/r$^{2}$ convolved with a Gaussian of width equal to the PSF
    (dashed line), although there is no physical motivation for
    this. The fact that the surface brightness rises towards A and the
    quasar suggests that both the active galaxies are sources of the
    photoionization of the bridge. } 
\label{qsodist}
\end{figure}

 The only emission line detected from the gas in the bridge is
  Ly$\alpha$, placing strong constraints on the ratio of
  CIV/Ly$\alpha$ and HeII/Ly$\alpha$. The lack of HeII suggests a
  relatively low ionization parameter, unlike that seen in gas
  coincident with jets in radio galaxies, whereas the lack of CIV
  emission suggests the gas is not hot enough to ionize CIV. Although
  gas near AGN typically has a high ionization parameter, low
  ionization parameter gas may arise at greater distances from the AGN
  where the photon field is more diffuse, particularly if the ionizing
  photons have already passed through a cloudy medium
  \citep[e.g.][]{Arrigoni-Battaia15}. The majority of ionizing photons
  exciting the bridge emission probably emanate from the quasar and
  the AGN within companion A as the Ly$\alpha$ surface brightness
  rises towards both (Fig.~\ref{qsodist}), although the exact form of
  the surface brightness profile will depend on the optical thickness
  and density distribution of the gas in the bridge. The large line
  width of Ly$\alpha$ suggests the bridge emission is not due to
  gravitational cooling, although gravitational cooling would produce
  little or no HeII and CIV \citep{Yang06}, and the Ly$\alpha$
  emission is unlikely to arise from shocks as this would result in a
  higher fraction of CIV compared to Ly$\alpha$ than that seen here
  \citep{Heckman91b}. The line ratios are consistent with
  photoionization from stars \citep{Prescott09}, although the
  Ly$\alpha$ equivalent width limit of the bridge is too high for the
  ionization source to be solely stellar (the equivalent width limit
  only allows for ionization from massive, young O stars and such a
  strong star burst would be a very short lived phenomenon making this
  an unlikely scenario). However, a more typical stellar population
  could provide some fraction of the ionizing photons along with
  ionizing photons from the two AGNs. Alternatively, the observed line
  ratios could just be an effect of the resonant properties of
  Ly$\alpha$ whereby Ly$\alpha$ will resonantly scatter out to larger
  distances from the quasar than non-resonant HeII
  \citep{Arrigoni-Battaia14}. However, this may not be a dominant
  effect as the photons that resonantly scatter are likely to escape
  the system on much smaller scales
  \citep{Arrigoni-Battaia15,Verhamme06,Cantalupo05}. 

 The double peaked Ly$\alpha$ profile of companion A, the quasar
  and the bridge (Fig.~\ref{spectra_Qzoom}) may be indicative of a
  surrounding absorbing large neutral gas cloud or halo, as is
  commonly found around other quasars \citep[e.g.][at similar
    redshifts]{Hennawi06,Prochaska13}. The absorption in all three
  components is at the same wavelength and goes nearly completely
  black implying that the absorbing material is separate to the
  emitting material rather than mixed in with it. This absorbing
  neutral halo has a radius $>50$ kpc centered on the quasar and hence
  a mass of at least $\sim$2 x 10$^3$ M$_{\odot}$ (using the column
  density estimate above). The column density of neutral material in
  this halo could be significantly higher than estimated above if in
  addition to the low column density material there are regions of
  high column density gas with lower covering fractions. These higher
  density clumps would likely emit Ly$\alpha$ photons given that they
  are probably optically thick clouds within the ionizing field of the
  quasar and the companion AGN. However, we do not detect a symmetric
  Ly$\alpha$ halo centered on the quasar, as opposed to the asymmetric
  bridge emission, suggesting the mass in higher density clumps is
  significantly less than that in the bridge. In terms of Ly$\alpha$
  absorption, the quasar's environment is consistent with the typical
  environment of high redshift quasars characterised by
  \citet{Prochaska13}. For the bridge and A to show similar absorption
  they would most likely have to be observed through a significant
  fraction of the absorbing halo, of a comparable column density to
  that affecting the quasar line of sight.

 Although radiative transfer effects can also produce double
  peaked Ly$\alpha$ emission in optically thick gas clouds
  \citep{Cantalupo05,Kollmeier10}, it seems improbable that this is
  the case here as the absorption wavelength is the same across the
  entirety of the bridge, quasar and companion A whilst the underlying
  Ly$\alpha$ emission profile changes significantly. Radiative
  transfer effects are unlikely to produce a double-peaked profile
  along the line-of-sight to the quasar, a point-source, and so the
  profile here most likely arises from multi-phase gas that includes
  neutral gas clouds in front of the quasar. The fact that the
  absorption profile directly in front of the quasar is so similar to
  that across the rest of the system - there are no discontinuities in
  the absorption wavelength - is highly suggestive that the quasar
  line-of-sight is probing the same relative distribution of material
  as the line-of-sight to the bridge. Indeed we see no other
  absorption lines in the quasar spectrum indicating that this
  covering neutral gas must have a low column density, consistent with
  the low column density estimate derived earlier assuming the
  quasar's double-peaked profile arises from absorbing material rather
  than radiative transfer effects.  

 The PKS1614+051 system is similar to that of SDSSJ0841+3921 and
  its three companion AGN that are also embedded within an extended
  Ly$\alpha$ halo, although at slightly lower redshift
  \citep{Hennawi15}. However, there is no evidence of merging activity
  or interaction in the SDSSJ0841 system and the Ly$\alpha$ spectral
  profiles from the quasar and extended nebula are not double-peaked,
  suggesting there is just one component of (multi-phase) gas present,
  unlike PKS1614+051 that seems to have additional ISM-like material
  inbetween the galaxies. The neutral halo surrounding PKS1614+051 may
  well fluoresce in Ly$\alpha$ in a similar manner to SDSSJ0841+3921
  but at a lower surface brightness than our detection limit. 

 Although the picture we put forward here fits with all the data
  it is not necessarily the only interpretation. These results are
  achieved with only 2 hrs of exposure. A significantly longer
  observation with MUSE would increase the signal-to-noise on the
  extended Ly$\alpha$ emission line allowing the spatial and velocity
  structure of the bridge to be determined in more detail, and
  potentially detect other emission lines from the bridge, allowing a
  plausible model of the photoionization source(s) to be put
  forward. Additionally, observations of OII, OIII and H$\alpha$ with
  JWST/NIRSPEC in the future would allow the metallicity and
  ionization parameter of the gas to be determined, potentially
  constraining the origin of the gas.

 As noted above, recent work on the gaseous environments of
quasars, in particular the extent of surrounding neutral haloes, has
used background probes \citep[e.g.~a
  quasar,][]{Hennawi06,Prochaska09,Prochaska13b} to explore the extent
of neutral gas in the target quasar's environment. The potential
advantages of the approach used here  in addition to the use of a
background probe is that MUSE can allow a search for absorption
against the extended emission  on 10s kpc scales,
allowing the whole region to be explored rather than just a single
sight line. The probability of identifying a background probe at small
projected distance for a quasar decreases as (radius)$^{-2}$ so the
approach used here can also better sample the haloes at small
radii. Clearly there is potential for observing a sample of quasars
with extended Ly$\alpha$ emission with MUSE in order to explore the
small scale quasar environment. This is particularly important as
\citet{Prochaska13} find an anisotropic distribution of N$_{HI}$
absorbers around quasars, with more absorbing clouds than expected
found around the quasar than in its line of sight (the proximity
effect), that they attribute to the quasar emitting anisotropically -
clouds along its line of sight are photo-evaporated but those
transverse to it are exposed to less radiation and survive. We see the
Ly$\alpha$ absorption in PKS1614 across the whole of the quasar
  and companion system suggesting that this quasar is radiating
relatively isotropically. In addition, given that we see the bridge
and both AGNs, and the bridge must lie mainly in the plane of the sky
or its true extent becomes improbably large, the opening angle for AGN
emission in both A and the quasar must be large and any obscuring
torus small.

The profile of the Ly$\alpha$ absorption is symmetric around the
quasars systemic velocity suggesting it is not associated with any
inflow or outflow of material. If material was in-flowing or out-flowing
it would need to be continually replenished or destroyed, or be a
short lived phenomenon, as the amount of absorber material is
relatively small ($>$ 2 x 10$^{3}$ M$_{\odot}$). Instead we postulate
this neutral material sits spherically around the quasar within its
dark matter halo and (some of) its large velocity width comes from the
fact that these clouds are moving within the gravitational potential
of the halo. It is clear that MUSE, with its large IFU coverage,
provides an excellent dataset to probe the quasar's environment over
an extended region, including providing important constraints on the
properties of its halo gas.

\section{Conclusions}
We find that the $z=3.215$ quasar PKS1614+051, with its
companion active galaxy and connecting bridge of Ly$\alpha$ emitting
gas, is surrounded by three other Ly$\alpha$ emitting galaxies, two of
which are very close to or within the bridge of material. This
overdensity of four galaxies, all within 300 km s$^{-1}$ of each other
and within the virial radius of the quasar, will likely merge to form
a massive elliptical galaxy by $z\sim1$, perhaps residing within a
group or cluster of galaxies.

 The 50 kpc long bridge of Ly$\alpha$ emitting gas stretches
  between the quasar and companion galaxy A and also extends towards
  the two other nearby companion galaxies, suggesting that the gas was
  pulled out from the galaxies during a gravitational interaction
  between them and the quasar host. The emission line bridge is likely
  to be predominantly photoionized by the quasar and its AGN companion
  as its surface brightness rises towards them both, although the lack
  of CIV or HeII in the bridge suggests a relatively low ionization
  parameter. The quasar's Ly$\alpha$ line is strongly double-peaked,
  likely due to absorption by a neutral halo of gas at the systemic
  velocity of the quasar extending at least 50 kpc in radius with a
  mass of at least $\sim$2 x 10$^{3}$ M$_{\odot}$.  The Ly$\alpha$
  bridge is probably a separate phase to the much larger neutral halo
  surrounding the quasar, which it is embedded in, as the Ly$\alpha$
  absorption profile goes nearly completely black indicating virtually
  all of the Ly$\alpha$ photons at that wavelength have been scattered
  out of the line of sight. In addition, we must be looking at the
  system through a significant path length of neutral material to see
  similar absorption across the entire system. This picture of the
  quasar, bridge and companion galaxies embedded in a large neutral
  halo is consistent with the absorption systems commonly found around
  quasars using a background quasar as a probe by
  e.g.~\citet{Prochaska13}. The fact that we see the bridge and both
  of the AGNs, and that we see neutral material in both the line of
  sight to the quasar and transverse to it, suggests the quasar is
  emitting isotropically and any obscuring torus obstructs only a
  small solid angle as seen from the nucleus.

The large field of the MUSE IFU allows the exploration of the quasar
environments on a wide range of scales. The clustering of galaxies at
the quasar redshift can be studied simultaneously with the small-scale
exploration of the quasar's gaseous environment using Ly$\alpha$
absorption of faint sources in the immediate environment. This use of
self-absorption, rather than using a rare background quasar to probe a
single line of sight, is a potentially powerful technique for
exploring the close environments of quasars, given the prevalence of
extended Ly$\alpha$ emission.

This work demonstrates the effectiveness and efficiency of MUSE for
elucidating the observational details of galaxy interactions and
growth at high redshift. While a single pointing can explore the early
growth of clustering, minimal tiling of several MUSE pointings should
ideally sample the larger-scale protocluster environment at these
redshifts.

\section{Acknowledgments} 
KH acknowledges funding from STFC. Based on observations made with ESO
telescopes at the La Silla Paranal Observatory under programme ID
60.A-9323(A). We thank all the staff at Paranal Observatory for their
valuable support during the commissioning of MUSE.

\bibliographystyle{mn2e}
\bibliography{bibliography}

\end{document}